\documentstyle[aps,pra,epsfig]{revtex}

\def\E{{\rm e}}
\def\I{{\rm i}}
\def\D{{\rm d}}
\begin{document}
\draft

\title{Single-file Diffusion with Random Diffusion Constants}

\author{Claude Aslangul\footnote
  {{\bf e-mail:} aslangul@gps.jussieu.fr}}
\address{Groupe de Physique des Solides, Laboratoire associ\'e au
CNRS (UMR 75-88), \\ Universit\'es Paris 7 \&
Paris 6, Tour 23, Place Jussieu, 75251 Paris Cedex 05, France}

\date{\today}

\maketitle
\begin{abstract}
The single-file problem of $N$ particles in one spatial dimension is
analyzed, when each particle has a randomly distributed diffusion constant
$D$ sampled in a density $\rho(D)$. The averaged one-particle distributions
of the
edge particles and the asymptotic ($N\gg 1$) behaviours of their transport
coefficients
(anomalous velocity and diffusion constant) are strongly dependent on the
$D$-distribution law, broad or narrow. When $\rho$ is exponential, it is
shown that the average
one-particle front for the edge particles does not shrink when $N$ becomes
very large, as
contrasted to the pure (non-disordered) case. In addition, when $\rho$
is a broad law, the same occurs for the averaged front, which can
even have infinite mean and variance. On the other hand, it is shown
that the central particle, dynamically trapped by all others as it is,
follows a narrow distribution, which is a Gaussian (with a diffusion
constant scaling as $N^{-1}$) when the fractional moment $<D^{-1/2}>$
exists and is finite; otherwise ($\rho(D)\propto D^{\alpha-1}$,
$\alpha\le\case12$), this density is, far from the origin, a
stretched exponential with an exponent in the range $]0,\,2]$; then the
effective diffusion constant scales as $N^{-\beta}$,
with $\beta\,=\,1/(2\alpha)$.
\end{abstract}

\vspace*{0.6cm}

\section{Introduction}\label{intro}
\setcounter{equation}{0}
The single-file diffusion problem is encountered in various
fields (one-dimensional hopping conductivity \cite{rich},
ion transport in biological membranes \cite{nener,sack},
channelling in zeolithes \cite{kukla}). Generally speaking, this is
modelled by a set of $N$ diffusing particles on the line with
hard-core repulsion; due to such an interaction, any initial ordering
is preserved in the course of time and the particles can be, once for
all, labelled
$1,\,2,\,\ldots,\,N$ from left to right. As in ref \cite {aslan}, I will
here consider the case where, at the initial
time, these particles form a compact cluster centered at the origin; the
solution for an arbitrary initial condition has been given in ref \cite
{Rod98}, using the reflection principle. In addition, as contrasted to
the standard model in which all the particles have the same diffusion
constant $D$, it is here assumed that the particle $i$ has a random
$D_{i}$, chosen independently of all the others in a given distribution
law $\rho(D)$. The latter can always be written as follows:
\begin{equation}
	\rho(D)\, =\,\frac{1}{D_{0}}\,r\left(\frac{D}{D_{0}}\right)
\label{rhoD}
	\enspace,
\end{equation}
where $D_{0}$ denotes a specific value of the diffusion constant
({\it e. g.} the average value) and $r(\xi)$ a positive function
normalized to unity:
\begin{equation}
 	\int_{0}^{+\infty}\,r(\xi)\,\D \xi =\,1\label{norm}
	\enspace,
\end{equation}
On a physical level, a random diffusion constant $D$ can arise in
various ways. For instance, by Stokes law and Einstein relation, it can
result from a random radius; more directly, this also happens for
particles having random masses.

The main purpose of this paper is to analyze how the choice of the
distribution $r$ modifies the large-$N$ dependence of the transport
coefficients for one particle of the cluster and, more generally, to
find out the density probability of its coordinate in the presence
of random diffusion constants. More specifically, focus
will be given on edge particles and on the one which (for $N$ odd)
is at the middle of the cluster; for the ``pure'' case (all the
particles have the same $D$), asymptotic laws ($N \gg 1$)
have been given in ref. \cite {aslan}. Among other results, it was
shown that increasing $N$ yields a narrowing of the one
particle-probability distributions, as a result of the ``pressure''
exerted by other particles on a given one. This will be shown not to
be true in all cases when the $D$'s are randomly sampled. As a rule,
if the middle particle is most often insensitive to disorder --
provided that the average $<D^{-1/2}>$ exists and is finite -- it will
be shown that the distribution for the side particles strongly
depends on the latter, going from narrow to broad laws.

\section{Basic relations}\label{basrel}
\setcounter{equation}{0}
The diffusion equation here writes:
\begin{equation}
 	\frac{\partial}{\partial t}\,p(x_{1},x_{2},\ldots,x_{N};t)\,=\,
 	\sum_{n=1}^{N}\,D_{n}\,\frac{\partial^{2}}{\partial
 	x_{n}^{2}}\,p(x_{1},x_{2},\ldots,x_{N};t)\label{Diff}
 	\enspace.
\end{equation}
As a consequence of the hard-core repulsion and of the above-mentionned
initial condition,
the solution has the expression:
\begin{equation}
 	p(x_{1},x_{2},\ldots,x_{N};t)\,=\,
 	C_{N}\,\prod_{n=1}^{N}\frac{\E^{-x_{n}^{2}/(4D_{n}t)}}{\sqrt{4\pi
D_{n}t}}\ \prod_{n=1}^{N-1}
 	Y(x_{n+1}-x_{n})\label{PN}
 	\enspace
\end{equation}
In (\ref {PN}), $Y$ is the Heaviside unit step function ($Y(x)=1$ if $x>0$,
$0$
otherwise), $D_{n}$ denotes the diffusion constant of the $n^{\rm th}$
particle and $C_{N}$ is the normalization constant. Since the $D$'s are
assumed
to be statistically independent, the various averages can be readily performed
and the average $N$-particle front turns out to be:
\begin{equation}
 	<p(x_{1},x_{2},\ldots,x_{N};t)>\,=\,
 	N!\,\prod_{n=1}^{N}\,<G(x_{n},t)>\ \prod_{n=1}^{N-1}
 	Y(x_{n+1}-x_{n})\label{PNmoyG} \enspace,
\end{equation}
where $<G(x_{n},t)>$ is the average of the gaussian distribution with
the density $\rho(D)$:
\begin{equation}
 	<G(x_{n},t)>\,=\,\int_{0}^{+\infty}\,\D D\,\rho(D)
 	\frac{\E^{-x_{n}^{2}/(4Dt)}}{\sqrt{4\pi Dt}}\,\equiv\,
 	 \frac{1}{(4D_{0}t)^{1/2}}\,f(u_{n})
 	\label{gaussmoyen} \enspace
\end{equation}
where:
\begin{equation}
 	f(u)\,=\,\frac{1}{\pi^{1/2}}\,\int_{0}^{+\infty}\,\D \xi\,
 	\xi^{-1/2}\,r(\xi)\,{\rm e}^{-u^{2}/\xi}\hspace{10mm}
 	u_{n}\,=\,\frac{x_{n}}{(4D_{0}t)^{1/2}}
 	\label{gmoyscal} \enspace.
\end{equation}
$f$ is an even positive function normalized to $\frac{1}{2}$ on the
interval $[0,\,+\infty]$.

From (\ref{PNmoyG}), all the one-particle averaged fronts can be
formally obtained by integrating over all $x$'s but one:
\begin{equation}
 	<p^{(1)}_{n}(x_{n};t)>\,=\,
 	\left(\prod_{m=1,\,m\neq n}^{N}\int_{-\infty}^{+\infty}\D
x_{m}\right)\,
 	<p(x_{1},x_{2},\ldots,x_{N};t)>
 	\label{p1n} \enspace.
\end{equation}
One then finds:
\begin{equation}
 	<p_{n}^{(1)}(x;t)>\,=\,\frac{N!}{(n-1)!(N-n)!}
 	\left[\frac{1+I(u)}{2}\right]^{n-1}\,
 	\left[\frac{1-I(u)}{2}\right]^{N-n}\,\frac{1}{(4D_{0}t)^{1/2}}\,f(u)\,
 	\label{p1nexpl}\enspace,
\end{equation}
where $I(u)$ is defined by:
\begin{equation}
 	I(u)\,=\,2\,\int_{0}^{u}\,\D u'\,f(u')\,=\,\int_{0}^{+\infty}\,\D\xi\,
 	r(\xi)\,\Phi(u/\sqrt{\xi})\,\equiv\,<\Phi(u/\sqrt{\xi})>
 	\label{Iu}\enspace
\end{equation}
where $\Phi$ denotes the probability integral \cite {Gr:Ry}.
In (\ref {p1nexpl}), the two factors $[1\pm I(u)]/2$ represent the steric
effects on the $n^{\rm th}$ particle due to all other ones.
$I(u)$ is a non-decreasing odd function
such that $I(\pm \infty)=\pm 1$ (for the pure case, one simply has
$I(u)=\Phi(u))$.
For the edge particles ($n=1$ or $n=N$), $<p_{n}^{(1)}(x;t)>$ can be
given a form occurring also in the theory of extreme events \cite {gum},
namely:
\begin{equation}
 	<p_{1,\,N}^{(1)}(x;t)>\,=\,\pm\,\,\frac{1}{(4D_{0}t)^{1/2}}\,
 	\frac{{\rm d}}{{\rm d}u}\,P_{\pm}(u)\hspace{10mm}
 	P_{\pm}(u)\,=\,\left[\frac{1\pm I(u)}{2}\right]^{N}
 	\label{p1edge}\enspace.
\end{equation}
The $-$ sign (resp $+$) refers to the left particle ($n=1$) (resp.
right particle ($n=N$)). In any case, the following relation holds
true:
\begin{equation}
 	<p_{1}^{(1)}(x;t)>\,=\,<p_{N}^{(1)}(-x;t)>
 	\label{p1edgepN}\enspace.
\end{equation}
For $n$ arbitrary, (\ref{p1edge}) generalizes into:
\begin{equation}
 	<p_{n}^{(1)}(x;t)>\,=\,\frac{1}{(4D_{0}t)^{1/2}}\,C_{N}^{n}\,
 	\left[\frac{{\rm d}}{{\rm
d}u}\,\left(\frac{1+I(u)}{2}\right)^{n}\right]
 	\,\left(\frac{1-I(u)}{2}\right)^{N-n}
 	\label{p1nqcq}\enspace.
\end{equation}
For the particle which is at the center of the cluster ($N$ odd), one has:

\begin{equation}
	<p_{(N+1)/2}^{(1)}(x;t)>\,=\,\frac{1}{(4D_{0}t)^{1/2}}\,\frac{N!}
{\{[(N-1)/2]!\}^{2}}
 	\left[\frac{1-I^{2}(u)}{4}\right]^{(N-1)/2}\,f(u)\,
 	\label{p1middle}\enspace,
\end{equation}
or, equivalently, from (\ref{p1nqcq}):
\begin{equation}
<p_{(N+1)/2}^{(1)}(x;t)>\,=\,\frac{1}{(4D_{0}t)^{1/2}}\,C_{N}^{(N+1)/2}\,
\left[\frac{{\rm d}}{{\rm d}u}\,\left(\frac{1+I(u)}{2}\right)^{(N+1)/2}\right]
\,\left(\frac{1-I(u)}{2}\right)^{(N-1)/2}
\label{pmiddle1}\enspace.
\end{equation}

In the following sections, the above equations will be used with
different choices for the distribution $r(\xi)$. Exact asymptotic results for
large $N$ can be obtained due to the fact that,
for $N\gg 1$, the steric factors $\{\frac{1}{2}[1\pm
I(u)]\}^{n}$ display a rather sharp variation.

Generally speaking, the one-particle transport coefficients can be
found from the behaviour at small $k$ of the characteristic function
$\Pi_{n}(K)$:
\begin{equation}
 	\Pi_{n}(K)\,=\,\,\int_{-\infty}^{+\infty}\,\D x\,\E^{\I
kx}\,<p_{n}^{(1)}(x;t)>
 	\hspace{10mm}(K=k\sqrt{4D_{0}t}\,)\label{Pi1caract}\enspace.
\end{equation}
The transport properties of the $n^{\tiny\mbox {th}}$ particle subjected to
the random
field of all the other are essentially described by the two first
cumulants:
\begin{equation}
 	<x_{n}>(t)\,=\,V_{1/2,\,n}(N)\,\sqrt{t}
	\hspace{10mm}\Delta\,x^{2}_{n}\,\equiv\,
	<x^{2}_{n}>(t)-[<x_{n}>(t)]^{2}\,=\,2D_{n}(N)\,t\,\enspace.
\end{equation}
The above time-dependences come from the fact that the
only available length-scale is $(D_{0}t)^{1/2}$, so that
$<x^{\lambda}>\,\propto\,t^{\lambda/2}$ at all times and for any
$\lambda$. Thus, at any time, the drift is
always anomalous (no velocity) and the mean square displacement always
has a purely diffusive motion. The following relations are trivially
veridied:
\begin{equation}
	<x_{1}>(t)\,=\,-\,<x_{N}>(t)\hspace{10mm}
	\Delta\,x^{2}_{1}(t)\,=\,\Delta\,x^{2}_{N}(t)
 	\label{x1edgexN}\enspace.
\end{equation}

In addition to the one-particle transport coefficients, statistical
correlations between the particles can be analyzed. For definiteness,
focus will be given on the correlations between the two edge
particles, altogether contained in the two-body density:
\begin{equation}
 	<p^{(2)}_{1,\,N}(x_{1},x_{N})>\,=\,
 	\left(\prod_{m=2}^{N-1}\int_{-\infty}^{+\infty}\D x_{m}\right)\,
 	<p(x_{1},x_{2},\ldots,x_{N};t)>
 	\label{p21N} \enspace.
\end{equation}
Using (\ref{PNmoyG}) -- (\ref{gmoyscal}), one finds:
\begin{equation}
 	<p^{(2)}_{1,\,N}(x_{1},x_{N};\,t)>\,=\,\frac{1}{4D_{0}t}\,N(N-1)\,
 	\left[\frac{I(u_{N})-I(u_{1})}{2}\right]^{N-2}
 	\,f(u_{1})f(u_{N})\,Y(u_{N}-u_{1})
 	\label{p21Nexpl} \enspace.
\end{equation}



\section{Exponential distribution}\label{expon}
\setcounter{equation}{0}
As a first simple example, let us choose:
\begin{equation}
 	\rho(D)\,=\,\frac{1}{D_{0}}\,\E^{-D/D_{0}}
 	\label{rhoexp}\enspace.
\end{equation}
Here, $D_{0}$ is the expectation value of $D$. From (\ref {gmoyscal})
and (\ref{Iu}), one readily obtains:
\begin{equation}
 	f(u)\,=\,\E^{-2\mid u\mid}\hspace{10mm}I(u)\,=\,{\rm
sgn}u\,(1-\E^{-2\mid u\mid})
 	\label{frhoexp}\enspace.
\end{equation}
The reduced densities for the edge particles density
can now be explicitely calculated from (\ref{p1edge}) with:
\begin{equation}
 	P_{-}(u)\,=\,\left\{ \begin{array}{ll}
               2^{-N}\,\E^{-2Nu} & \mbox{if $u\,\ge\,0$}\\
               (1-\frac{1}{2}\,\E^{-2\mid u\mid})^{N}& \mbox{if $u\,\le\,0$}
    \end{array}\right.
 	\label{P1xp}\enspace
\end{equation}
and $P_{+}(u)=P_{-}(-u)$. From (\ref{P1xp}), the generating
functions $\Pi_{1,\,N}(K)$ (\ref {Pi1caract})
for the edge particles are found as the following:
\begin{equation}
 	\Pi_{1}(K)\,=\,1+\I K\,\left[\frac{1}{2^{N}(N-\I K)}\,+\,
 	\sum_{p=1}^{N}\,C_{N}^{p}\frac{(-1)^{p}}{p+\I K}\right]
 	\hspace{10mm}\Pi_{N}(K)\,=\,\Pi^{*}_{1}(K)\label{Pi1NCarexp}\enspace.
\end{equation}
Now, by expanding $\Pi_{1,\,N}$ in the vicinity of $K=0$, and by coming
back to
the $x$-variable, one obtains:
\begin{equation}
	<x_{N}>(t)\,=\,-\,<x_{1}>(t)\,=\,[\ln(N/2)+C\,
	+\,\mbox{O}(2^{-N})]\,\sqrt{D_{0}t}\,
 	\equiv\,V_{1/2,\,\mbox{\small edge}}(N)\,\sqrt{D_{0}t}
 	\label{xexp}\enspace
\end{equation}
where $C$ is the Euler's constant ($C= 0.577\ldots$). The mean square
displacements are:
\begin{equation}
	\Delta\,x^{2}_{N}\,=\,\Delta\,x^{2}_{1}\,=\,[\frac{\pi^{2}}{6}\,
	+\,\mbox{O}(2^{-N})]\,D_{0}t\,
 	\equiv\,2D_{\mbox{\small edge}}(N)\,t\label{Dx2exp}\enspace.
\end{equation}
As compared to the pure case, the large-$N$ behaviour of the
transport coefficients is frankly different. When all the particles
have the same diffusion constant, one has \cite{aslan}:
\begin{equation}
 	V_{1/2,\,\mbox{\small edge}}^{\mbox{\small (pure)}}(N)\,\propto\,(\ln
 	N)^{\case12}\hspace{10mm}
 	D_{\mbox{\small edge}}^{\mbox{\small (pure)}}(N)\,\propto\,(\ln N)^{-1}
 	\label{VDexp}\enspace.
\end{equation}

As contrasted, for exponentially distributed random $D$, the anomalous drift
coefficient increases like $\ln N$, whereas the
effective diffusion constant $D_{\mbox{\small edge}}(N)$ tends towards
a {\it finite} (non-vanishing)
value. Also note that the height of the maximum also saturates and
does not go to zero for $N$ infinite. As a whole, the effect of an
exponential disorder is rather subbtle: most particles
of the cluster have a rather small $D$, so that the pressure they exert
on the edge ones is smaller than in the pure case; on the other hand,
large $D$ forces the packet to more with a ``velocity'' which
increases more rapidly with $N$ than in the pure case. The interplay
of these facts produces a distribution having a finite height and
width at infinite $N$. This is illustrated on Fig. \ref{p1gexp}, which clearly
shows the width and height saturation when $N$ becomes very large, as
quantitatively described in
eq. (\ref{Dx2exp}). It is readily seen that the value of $<p_{1,\,N}^{(1)}>$
at its maximum is close to $(2/\E)\,\propto\,N^{0}$.

For $N\gg 1$, the abscissa of the edge particles are
approximately  distributed according to:
\begin{equation}
	<p_{N}^{(1)}(x;t)>\,\simeq\,\,Y(u)\,\frac{N}{(4D_{0}t)^{1/2}}\,
	\E^{-(N/2)\,\E^{-2u}}
 	\,\E^{-2u}\hspace{5mm}<p_{1}^{(1)}(x;t)>\,=\,<p_{N}^{(1)}(-x;t)>
 	\label{p1Nas}\enspace.
\end{equation}
These functions are exponentially small ($\sim \E^{-N}$) near $\mid
u\mid \,=0$ and display a plain exponential decay ($\sim \E^{-2\mid
u\mid}$) for $\mid u\mid \,\gg \,\ln[(N/2)^{1/2}]$. $<p_{N}^{(1)}>$ is
maximum for $u\,=\,u_{\small\mbox{max}}\,=\,\,\case 12\,\ln (N/2)$; this
shows that
the corresponding $x_{\small\mbox{max}}$ coincides with $\pm <x_{N}>$
(see (\ref{xexp})). The approximate
expressions (\ref{p1Nas}) are hardly distinguishable from their exact
counterparts as soon as $N$ is greater than a rather small number
($\simeq 50$) and can be rewritten in terms of the proper shifted (but
here not rescaled) variable $X$:
\begin{equation}
	<p_{N}^{(1)}(x;t)>\,\simeq\,\,Y(x)\,\frac{1}{(D_{0}t)^{1/2}}\,\,
	\E^{-\,\E^{-X}}\,\E^{-\,X}
	\hspace{8mm}X\,=\,\frac{x\,-\,x_{\small\mbox{max}}}{(D_{0}t)^{1/2}}
\hspace{8mm}x_{\small\mbox{max}}=(D_{0}t)^{1/2}\,\ln(N/2)\label{p1NasX}\enspace.
\end{equation}

The statistical correlations between the two edge particles are most
simply measured by the correlator $C_{1N}$:
\begin{equation}
 	C_{1N}\,=\,<x_{1}\,x_{N}>\,-\,<x_{1}>\,<x_{N}>
 	\label{correxpo}\enspace
\end{equation}
which, due to scaling in space, varies linearly in time. $C_{1N}$ can be
found by using (\ref{p21Nexpl}) and (\ref{frhoexp}); for $N \gg 1$,
a little algebra yields:
\begin{equation}
 	C_{1N}\,=\,4\,D_{0}t\,\left[\ln^{2} (N/2)+\mbox{O}(\ln N)\right]
 	\label{correxp}\enspace.
\end{equation}
By (\ref{Dx2exp}), this implies that the normalized ratio
$C_{1N}/\Delta\,x^{2}_{1,N}$ -- which is constant in time -- has a
logarithmic increase with $N$:
\begin{equation}
	\frac{C_{1N}}{\Delta\,x^{2}_{1,N}}\,\simeq\,\frac{6}{\pi^{2}}\,\ln^{2} (N/2)
	\
label{correxpnorm}\enspace.
\end{equation}
Although the numerator and the denominator of the ratio have
separately different behaviours as compared to the pure case, the
ratio still has a $\sim (\ln N)^{2}$ increase with $N$, exactly as in this
latter case\cite{aslan}.

For the particle located at the middle of the cluster, the expression
(\ref {pmiddle1}) writes:
\begin{equation}
<p_{(N+1)/2}^{(1)}(x;t)>\,=\,\frac{1}{(4D_{0}t)^{1/2}}\,\frac{1}{2}\,
C_{N}^{\,(N+1)/2}\,
\left(1-\frac{1}{2}\,\E^{-2\mid u\mid}\right)^{(N-1)/2}\,
\E^{-(N+1) \mid u\mid}
\label{pmiddle1exp}\enspace.
\end{equation}
From (\ref {pmiddle1exp}), the following large-$N$ expression of
the mean square dispersion is obtained:
\begin{equation}
 	\Delta x_{(N+1)/2}^{2}(x;t)\,=\,\frac{1}{2}D_{0}t\,C_{N}^{(N+1)/2}
	\,\left[B\left(\frac{N+1}{2},\frac{N+1}{2}\right)-4B\left(\frac{N+3}{2},
	\frac
{N+3}{2}\right)\right]
 	\label{Dmiddleexp}\enspace,
\end{equation}
where $B$ denotes the Euler beta-function \cite {Gr:Ry}. Expansion for
large-$N$ yields:
\begin{equation}
 	\Delta x_{(N+1)/2}^{2}(x;t)\,\simeq\,\frac{1}{N}D_{0}t
 	\label{Dmiddle1exp}\enspace
\end{equation}
and provides the asymptotic dependence of the diffusion constant:
\begin{equation}
 	D_{\mbox{\small middle}}(N)\,\simeq\,\frac{1}{2N}\,D_{0}
 	\label{Dmiddlexp}\enspace.
\end{equation}
This behaviour is the same as for the pure case\cite{aslan}. Thus,
for the exponential distribution, the transport coefficient of
the central particle is essentially unaltered by disorder -- as shown
below this is also true for any $r$ density having $<D^{-1/2}>$ finite. The
asymptotic density of the middle particle is obtained from (\ref
{pmiddle1exp}) as the following:
\begin{equation}
 	<p_{(N+1)/2}^{(1)}(x;t)>\,\simeq\,\left(\frac{N}{2\pi
D_{0}t}\right)^{1/2}\,
 	\E^{-2(\mid u\mid\,+\,N\,u^{2})}
 	\label{pmidas}\enspace.
 \end{equation}
This function is normalized to unity up to $N^{-1}$ terms. Apart from
the cusp at $u=0$ and outside the region $\mid u\mid \,\le \,1/N$, this is
essentially a gaussian distribution, exactly as in the pure case. For
large $N$, the cusp can even be ignored and the distribution looks
like a Gaussian everywhere.

In fact, it can now be stated the conditions for the stochastic dynamics
of the central particle to be unaffected by disorder; such a
particle is stymied in some way due to other particles and it can be
suspected that strong steric effects most often dominate, except perhaps when
the distribution $\rho(D)$ gives a high probability to find very
small $D$.
The analysis goes as follows. For $N \gg 1$, it is readily seen that
the distribution (\ref{p1middle}) assumes non-negligible values only
for $\mid u \,\mid \ll 1$. This in turn allows to replace $I(u)$ by its
small-$u$ expansion; from (\ref{Iu}) and (\ref{gaussmoyen}), $f(0)$
exists and is finite provided
that $\rho(D)$ is bounded by $D^{-\beta}$ ($\beta <\,\case12$) for
$D\rightarrow 0$.
This means that when $\rho$ diverges at small $D$, this divergence is
not too severe, which implies that finding a small $D$ has not such a great
probability. With this assumption, one has:
\begin{equation}
	I(u)\,\simeq\,2u\,f(0)\hspace{10mm}f(0)\,=\,\frac{1}{\pi}\,\int_{0}^{+\infty}\,
\D \xi\, \xi^{-1/2}\,r(\xi)
\,\equiv\,\frac{1}{\pi}\,<\left(\frac{D_{0}}{D}\right)^{1/2}>\label{Ismallu}
\enspace.
\end{equation}
Using Stirling formula, and expanding
$\E\,^{[(N-1)/2]\,\ln[1-4u^{2}f^{2}(0)]}$, one eventually gets the
asymptotic approximation:
\begin{equation}
 	<p_{(N+1)/2}^{(1)}(x;t)>\,\simeq\,\left(\frac{N}{2\pi\,
 	D_{0}t}\right)^{1/2}\,f(0)\,\E^{-2N\,f^{2}(0)\,u^{2}}
 	\label{pmidasfin}\enspace.
\end{equation}
Note that the approximation automatically generates a properly normalized
density. So, for any $\rho(D)$ such that $\lim_{D\rightarrow
0}[D^{1/2}\rho(D)]\,=\,0$, the middle particle is essentially distributed
according a normal law, the width of which decreasing as $N^{-1/2}$,
up to unessential numerical factors. The expression (\ref {pmidasfin})
can be written as a universal law in term of the proper scaled
variable $X$:
\begin{equation}
 	<p_{(N+1)/2}^{(1)}(x;t)>\,\simeq\,\left(\frac{N}{\pi
t}\right)^{1/2}\,<D^{-1/2}>
 	\,\frac{\E^{-\,X^{2}/2}}{(2\pi)^{1/2}}\,\hspace{10mm}
 	X\,=\,x\,\left(\frac{N}{\pi t}\right)^{1/2}\,<D^{-1/2}>
 	\label{pmidasuniv}\enspace,
\end{equation}
expressing as a whole the irrelevance of details about the $D$
distributions for the central particle when $\rho(D)$ is bounded near
$D=0$ as stated above, which entails that the fractional moment
$<D^{-1/2}>$ exists and is finite. The opposite case is treated in the next
section which deals with the Gamma-distribution.

%


\section{Gamma distribution}\label{gamma}
\setcounter{equation}{0}
As another example -- which in fact contains the exponential
distribution as a particular case -- let us choose a Gamma
distribution:
\begin{equation}
	\rho(D)\, =\,\frac{1}{\Gamma(\alpha)\,D_{0}^{\alpha}}\,D^{\alpha -
1}\,\E^{-D/D_{0}}
	\hspace{10mm}(\alpha > 0) \label{rhoDgamma}
	\enspace.
\end{equation}
With (\ref {rhoDgamma}), the expectation values of $D$ and $D^{2}$
are respectively equal to $\alpha D_{0}$ and
$\alpha(\alpha+1)D_{0}^{2}$. The exponential distribution is
recovered by setting $\alpha=1$, whereas the pure case can be
obtained by taking the limit $\alpha \rightarrow \infty$, $D_{0}\rightarrow
0$, $\alpha D_{0}\,=\,\mbox{const}$. The $f$ function (\ref{gmoyscal}) is:
\begin{equation}
	f(u)\, =\,\frac{2}{\sqrt{\pi}\,\Gamma(\alpha)}\,\mid u\mid^{\alpha -
	\case12}\,K_{\alpha - \case12}(2\mid u\mid)
	\label{fDgamma}
	\enspace
\end{equation}
where $K_{\alpha - \case12}$ is the Bessel function of imaginary
argument\cite {Gr:Ry}. $f(0)$ exists and is finite if
$\alpha\,>\,\case12$:
\begin{equation}
	f(0)\,=\,\frac{1}{\sqrt{\pi}}\,\frac{\Gamma(\alpha+\case12)}{(\alpha-\case12)\,
\Gamma(\alpha)}
	\hspace{10mm}(\alpha\,>\,\frac{1}{2})\label{fDgammazero}
	\enspace
\end{equation}
The large-$N$ approximation of the
one-particle densities can be obtained along the same lines as in the
previous section, by using the approximate expression:
\begin{equation}
	I(u)\, \simeq\,1\,-\,\frac{1}{\Gamma(\alpha)}\,\mid u\mid^{\alpha -
1}\,\E^{-2\mid u\mid}
	\hspace{5mm}(\mid u\mid \gg 1)
	\label{Iugammaapp}
	\enspace.
\end{equation}

Let us consider first the right particle; one has:
\begin{equation}
  	<p_{N}^{(1)}(x;t)>\,\simeq\,\,Y(u)\,\frac{1}{(4D_{0}t)^{1/2}}\,
  	\frac{N}{\Gamma(\alpha)}\,\E^{-\frac{N}{2\Gamma(\alpha)}\,
  	u^{\alpha-1}\,\E^{-2u}}\,u^{\alpha -1}\,\E^{-2u}
 	\label{p1Ngammas}\enspace.
\end{equation}
When $\alpha\,<\, 1$, {\it i. e.} when $\rho$ diverges at small $D$,
$<p_{N}^{(1)}>$ is exponentially small near $x=0$. On the contrary,
for $\alpha\, > \,1$, $<p_{N}^{(1)}>$ strictly vanishes as
$u^{\alpha-1}$, but this behaviour is realized on a very small
interval, namely $u\,<\,N^{-1/(\alpha-1)}$, beyond which $<p_{N}^{(1)}>$
remains exponentially small. On
the other hand, at large $u$, $<p_{N}^{(1)}>$ has essentially an
exponential decay, so that all the average values $<x_{N}^{m}>$ exist
and are finite. $<p_{N}^{(1)}>$ is maximum at $u_{\mbox{\small max}}$:
\begin{equation}
  	u_{\mbox{\small
max}}\,\simeq\,\frac{1}{2}\,\ln\frac{N}{2\Gamma(\alpha)}
 	\label{umaxgamma}\enspace.
\end{equation}
Again, $x_{\mbox{\small max}}\,=\,(4D_{0}t)^{1/2}\,u_{\mbox{\small
max}}$ is the relevant rescaled variable, since (\ref{p1Ngammas}) can
be rewritten as follows:
\begin{equation}
  	<p_{N}^{(1)}(x;t)>\,\simeq\,\,Y(u)\,\frac{1}{(D_{0}t)^{1/2}}\,
  	\E^{-\,(u/u_{\mbox{\tiny max}})^{\alpha-1}\,\E^{-2(u/u_{\mbox{\tiny
max}})}}\,
  	(u/u_{\mbox{\small max}})^{\alpha -1}\,
  	\E^{-2(u/u_{\mbox{\tiny max}})}
 	\label{p1NgammasX}\enspace.
\end{equation}
This readily gives $<p_{N}^{(1)}>_{\mbox{\small max}}\,\propto\,N^{0}$ which
in turns allows to guess that, as for the exponential case, one has
$<x_{N}>\,\propto\,\ln N$ and $<\Delta
x_{1,\,N}^{2}>\,\propto\,N^{0}$; it thus turns out that detailed
calculations for $\alpha \neq 1$ are indeed of little interest.
Up to unessential numerical factors, the large-$N$ asymptotic
laws all belong to the same class for Gamma-distributed diffusion
constants, the behaviours being rather easily obtained by
considering the simpler exponential case $\alpha\,=\,1$; the same
obviously holds for statistical correlations between edge particles.

For the central particle, one has to distinguish the two cases
$\alpha\,>\,\case12$ and $\alpha\,<\,\case12$. Due to (\ref
{fDgammazero}), the first one is described by the general expression
(\ref{pmidasuniv}). On the contrary, for $\alpha\,<\,\case12$, one has:
\begin{equation}
  	f(u)\,\simeq\,\,\frac{2^{2\alpha-1}}{\Gamma(2\alpha)\,\cos\alpha\pi}\,
  	\mid u\mid ^{2\alpha-1}\hspace{10mm}(\mid u\mid \, \ll \,1)
 	\label{fgamupetit}\enspace.
\end{equation}
This in turns entails that:
\begin{equation}
  	F(u)\,\simeq\,\,\frac{2^{2\alpha}}{\Gamma(2\alpha+1)\,\cos\alpha\pi}\,
  	\mid u\mid ^{2\alpha}\hspace{10mm}(\mid u\mid \, \ll \,1)
 	\label{Fgamupetit}\enspace.
\end{equation}
As a consequence, one obtains:
\begin{equation}
  	<p_{(N+1)/2}^{(1)}(x;t)>\,\simeq\,\frac{\alpha\,A\,}{(D_{0}t)^{1/2}}\,
  	\sqrt{\frac{2N}{\pi}}\,\E^{-(N/2)\,A^{2}\,\mid u\mid ^{4\alpha}}\,\mid
  	u\mid ^{2\alpha -1}
	\hspace{10mm}A\,=\,\frac{1}{\sqrt{\pi}}\,
	\frac{\Gamma(\frac{1}{2}-\alpha)}{\alpha\Gamma(\alpha)}
 	\label{p1midgamas}\enspace.
\end{equation}

This shows that when the probability is very high to find quite small
diffusion constants, the distribution for the central particle
diverges at $x\,=\,0$ (but is clearly integrable) and is essentially a
stretched exponential when $\mid u\mid$ is large:
\begin{equation}
 	<p_{(N+1)/2}^{(1)}(x;t)>\,\propto\,\left\{ \begin{array}{ll}
               \mid u\mid^{-(1-2\alpha)} \hspace{5mm}& \mbox{if \,\,$\mid
u\mid\,\ll\,1$}\\
               \E^{-\,C\mid u\mid ^{4\alpha}}\hspace{5mm}& \mbox{if \,\,$\mid
               u\mid\,\gg\,1$}
    \end{array}\right.
 	\label{p1midgasbeh}\enspace.
\end{equation}
For $\alpha=\case12$, $f(u)$ is proportionnal to the Bessel function
$K_{0}$ and the divergence is logarithmic. In this case, one has:
\begin{equation}
  	<p_{(N+1)/2}^{(1)}(x;t)>\,\simeq\,-\,\frac{1}{(D_{0}t)^{1/2}}\,
  	\sqrt{\frac{2N}{\pi^{3}}}\,\E^{-(16N/\pi^{2})\,u^{2}\,\ln\mid
u\mid}\,\ln \mid u\mid
 	\label{p1midgamas12}\enspace
\end{equation}
which entails:
\begin{equation}
 	<p_{(N+1)/2}^{(1)}(x;t)>\,\propto\,\left\{ \begin{array}{ll}
               -\,\ln\mid u\mid \hspace{5mm}& \mbox{if \,\,$\mid
u\mid\,\ll\,1$}\\
               \E^{-\,C'u ^{2}}\hspace{5mm}& \mbox{if \,\,$\mid
               u\mid\,\gg\,1$}
    \end{array}\hspace{5mm}(\alpha\,=\,\case12)\right.
 	\label{p1midgas12}\enspace.
\end{equation}
The divergence at $x=0$ comes from the interplay of pressure and of the
frequent occurrence of central particle with a very small $D$. From (\ref
{p1midgamas}), one finds:
\begin{equation}
  	\Delta x_{(N+1)/2}^{2}\,\simeq\,8\,
  	\left[\frac{\Gamma(\alpha
+1)}{\Gamma(\frac{1}{2}-\alpha)}\right]^{1/\alpha}\,
	\frac{\Gamma(\frac{1}{\alpha})}{\Gamma(\frac{1}{2\alpha})}\,\left(\frac{2N}{\pi
}\right)^{-\frac{1}{2\alpha}}\,D_{0}t
 	\label{x2gamas}\enspace.
\end{equation}

Summing up, the effective diffusion constant for the central particle
and for the Gamma distribution scales with $N$ as follows:
\begin{equation}
 	D_{\mbox{\small middle}}\,\propto\,N^{-\beta}
 	\label{expogamma}\enspace
\end{equation}
with:
\begin{equation}
 	\beta\,=\,\left\{ \begin{array}{ll}
               \,\,1\hspace{5mm}& \mbox{if \,\,$\alpha\,\ge\,1/2$}\\
               \frac{1}{2\alpha}\hspace{5mm}& \mbox{if
\,\,$0\,<\,\alpha\,\le\,1/2$}
    \end{array}\right.
 	\label{edxpgam}\enspace.
\end{equation}


\section{Broad distributions}\label{broad}
\setcounter{equation}{0}
The two previously examples discard the interesting case in which the
diffusion constants are sampled in a broad law, possibly devoid of
usual first few moments (mean and variance). Obviously enough, such a
case is interesting since the average front for the edge particles are
expected to be strongly asymmetric and also quite diffuse, due to the
pressure exerted by inner
particles and the possibility of rather high diffusion contants. This fact
even opens the
possibility for the edge particules to be also distributed according  a
broad law, devoid of mean and variance. Generally speaking, I will now
consider
the case:
\begin{equation}
	\rho(D)\, =\,\mu\,D_{0}^{\mu}\,Y(D-D_{0})\,D^{-(\mu +1)}\hspace{5mm}
	\Longleftrightarrow\hspace{5mm} r(\xi)\,=\,Y(\xi -1)\,\xi^{-(\mu +1)}
	\hspace{10mm}(\mu > 0)
	\label{rhoDbroad}
	\enspace.
\end{equation}
All the moments $<D^{k}>$ diverge for $k\ge \mu$.

With the distribution (\ref {rhoDbroad}), the $f$ function (\ref
{gaussmoyen}) can be expressed in terms of the incomplete $\gamma$
function \cite{Gr:Ry}:
\begin{equation}
	f(u)\, =\,\frac{\mu}{\pi^{1/2}}\,u^{-(2\mu +1)}\,\gamma(\mu
+\case12,\,u^{2})
	\label{fDbroad}
	\enspace.
\end{equation}
Again, I am interested in the large-$N$ limit, in which case the right
particle density (\ref{p1edge}) assumes non-negligible values for $u$ of
the order of
or greater than $u_{0}$ defined as:
\begin{equation}
	\left[\frac{1+I(u_{0})}{2}\right]^{N}\, =\,\frac{1}{2}\hspace{5mm}
	\Longleftrightarrow\hspace{5mm}u_{0}\,\simeq\,\left[\frac{N}{\pi^{1/2}\,\ln 	4
}\,\Gamma(\mu+\case12)\right]^{\,1/(2\mu)}
	\label{defu0}
	\enspace.
\end{equation}
For genuine broad laws ($\mu$ small), $u_{0}$ is much greater than 1 even
for moderately
large values of $N$. This allows to state again that the large-$u$ expansion
of $I(u)$ is only relevant, and to substitute everywhere the approximate
expression:
\begin{equation}
	I(u)\,\simeq\,1\,-\,\frac{1}{\pi^{1/2}}\,\Gamma(\mu+\case12)
	\,u^{\,-2\mu}\hspace{5mm}(u\gg 1)
	\label{Iuapprox}
	\enspace.
\end{equation}
Injecting this in (\ref{p1edge}) yields the large-$N$ approximation,
valid for $u\,>\,0$:
\begin{equation}
	<p^{(1)}_{N}(x;t)>\,\simeq\,N\,\frac{\mu}{\pi^{1/2}}\,
	\E^{-\,N\,\Gamma(\mu+\case12)/(2\pi^{1/2}\,u^{2\mu})}\,u^{-(2\mu+1)}\,
	\gamma(\mu+\case12,u^{2})
	\label{pNedgebroadas}
	\enspace,
\end{equation}
it being understood that $<p^{(1)}_{N}>$ is exponentially small for $u<0$
and can be considered as identically vanishing for $u<0$; note that
the expression
(\ref{pNedgebroadas}) goes toward zero extremely rapidely when
$u\,\rightarrow\,0$, namely:
\begin{equation}
	<p^{(1)}_{N}>\,\sim\,\E^{-\mbox{\small Cst}\,N/u^{2\mu}}
	\label{pNedgebrsmallu}
	\enspace.
\end{equation}
On the other hand, this distribution is indeed a broad law in the wide
sense, since
(\ref{pNedgebroadas}) displays for any $\mu$ a power-law behaviour at
(very) large $u$:
\begin{equation}
	<p^{(1)}_{N}(x;t)>\,\simeq\,N\,\frac{\mu}{\pi^{1/2}}\,
	\Gamma(\mu+\case12)\,u^{-(2\mu+1)}
	\hspace{5mm}(u\gg N^{1/(2\mu)})
	\label{pNedgebroadasdev}
	\enspace.
\end{equation}
Eqs. (\ref{pNedgebrsmallu}) and (\ref{pNedgebroadasdev}) show that the
average front for the right particle is strongly
asymmetric, displaying a rather steep increase on the left (towards the
inner part of the cluster) and a very slow decrease on the other
side, towards the free part of space. The result (\ref{pNedgebroadasdev})
entails that the moment $<u^{m}>$ exists and is finite
only if the following inequality is satisfied:
\begin{equation}
	m\,<\,2\,\mu
	\label{momexist}
	\enspace.
\end{equation}
Thus, for $\mu\le \case12$, the expectation value of $x$ (as well as all
higher moments) is infinite. The one-particle distribution (plotted in
Fig. \ref {p1dbroad} for a few values of $N$) can
nevertheless be characterized by the value $u_{\mbox{\small max}}$ giving its
maximum value, which turns out to coincide with $u_{0}$ (see
(\ref{defu0})), except for numerical factors. It is readily verified that,
accordingly, the value
of  $<p^{(1)}_{1,\,N}>$ at its maximum is $\,\propto\,N^{-\,1/(2\mu)}$.
As a whole, the maximum of $<p^{(1)}_{N}(x;t)>$ moves in time as the
following:
\begin{equation}
	x_{\mbox{\small
max}}(t)\,\simeq\,2\,\left[\frac{N}{\pi^{1/2}}\,\Gamma(\mu+\case12)\,
	\frac{\mu}{2\mu +1}\right]^{\,1/(2\mu)}\,(D_{0}t)^{1/2}
	\label{xmax}
	\enspace.
\end{equation}
$x_{\mbox{\small max}}$ is here the relevant rescaled coordinate;
agreement is quite good with the exact results (see Fig.
\ref{p1dbroad}).

Note that for $\mu \rightarrow\,\infty$, (\ref{pNedgebroadasdev}) shows that
$<p^{(1)}_{N}>$ decreases faster than any arbitrary power of $u^{2}$,
in agreement with \cite{aslan}. Also note that no obvious measure of
correlations here exists since for $\mu < 1$ the mean square
dispersions are infinite.

\section{Summary and Conclusions}\label{sumcon}
\setcounter{equation}{0}
For the single-file diffusion problem with random diffusion constants, the
asymptotic laws (large $N$) giving the transport coefficients and
the averaged one-particle densities have been derived.

Generally speaking, the asymptotic distribution of the central
particle is most often gaussian (see (\ref{pmidasfin})), as a result of the
fact that steric
effects nearly always dominate; when this is the case, the dynamical
trapping of
this particle is insensitive to details describing the distribution
$\rho(D)$
of the diffusion constants. On the other hand, when $\rho(D)$
diverges as $D^{-1/2}$ or faster at small $D$ ($\rho(D)\propto
D^{\alpha-1}$, $\alpha\le\case12$), the distribution of the
central particle is no more a normal one: at small $x$, it itself
goes to infinity as $x^{-(1-2\alpha)}$ and is a stretched exponential
$\E^{-Cx^{4\alpha}}$ at large $x$
(see (\ref{p1midgamas})). The effective diffusion
constant scales as $N^{-1/(2\alpha)}$ for $\alpha\,<\,\case12$ and as
$N^{-1}$ otherwise, showing that in the $N$-infinite
limit, as in the pure case, the motion indeed becomes subdiffusive in
agreement with Harris
result \cite{harris}.

On the contrary, the stochastic dynamics of the
edge particles is strongly dependent upon the nature (narrow or
broad) of the diffusion constants probability density $\rho(D)$. When the
latter
is narrow ({\it i. e.} exponential), the effective diffusion constant
of edge particles tends toward a {\it finite} constant when $N$
becomes infinite; the average
coordinate increases with $N$ as $\ln N$, faster than for the pure
case. When the $\rho(D)$ is broad ({\it i. e.} behaves as a power
law $D^{-(\mu+1)}$ at large $D$), the edge particles are themselves
distributed according a broad law, devoid of mean and variance when
$\mu$ is smaller than $\case12$. The distribution of the coordinate,
strongly asymmetric (see Fig. \ref {p1dbroad}),
can nevertheless be characterized by the abscissa of its maximum (\ref
{xmax}); the prefactor of the latter displays a $N^{1/(2\mu)}$
scaling {\it i. e.} rapidly increases with $N$ when $\mu$ is small.
This comes from the fact that there is a high probability to find inner
particles with a
large diffusion constant which, in turns, gives rise to a high
``pressure'' exerted by the core of the cluster on its ``surface''.

In all cases, the averaged one-particle asymptotic density has been found; it
can be written under a general form displaying the basic ingredients
of the problem. For the right particle, one has:
\begin{equation}
	<p_{N}^{(1)}(x;\,t)>\,\simeq\,N\,\mbox{exp}\left[-\,\frac{N}{2}\,
	<1-\Phi[x/(4Dt)^{1/2}]>\right]\,\,<G(x,\,t)>
	\label{pNgeneral}
	\enspace.
\end{equation}
where $<\ldots>$ denotes averaging with the distribution $\rho(D)$ of the
diffusion constants, $\Phi$ is the probability integral and $G$ is the
gaussian distribution. For the central particle ($N$ odd), one has:
\begin{equation}
	<p_{(N+1)/2}^{(1)}(x;\,t)>\,\simeq\,\left(\frac{N}{2\pi^{2}t}\right)^{1/2}\,<D^
{-1/2}>\,
	\mbox{exp}\left[-\,N<D^{-1/2}>\frac{x^{2}}{2\pi t}\right]\,
	\label{pcentgen}
	\enspace,
\end{equation}
provided that the fractional moment $<D^{-1/2}>$ exists and is
finite. Otherwise, when $\rho\,\propto D^{\alpha-1}$ at small $D$
with $\alpha\,\le\,\case12$, $<p_{(N+1)/2}^{(1)}>$ is of the form (see
(\ref {p1midgamas}) for details):
\begin{equation}
	<p_{(N+1)/2}^{(1)}(x;\,t)>\,\simeq\,C\,\mid x\mid^{-(1-2\alpha)}\,
	\mbox{exp}\left[-\,C'\mid x\mid^{4\alpha}\right]
	\label{pcentgen12}
	\enspace.
\end{equation}
At large $x$, this gives a stretched exponential with an exponent in
the range $]0,\,2]$, whereas the density diverges at $x=0$ as a
power-law ($\propto x^{-(2\alpha -1)}$); for the particular case
$\alpha=\case12$, this singularity becomes logarithmic.


\newpage
{\bf Figure Captions}
\begin{enumerate}
\item{}
\label{p1gexp}
Exact averaged front $<p_{1}^{(1)}(x;t)>$ for the left particle with
exponentially
distributed diffusion constants; the abscissa is the dimensionless
variable $u=x/(4D_{0}t)^{1/2}$. Each curve is labelled by the number
of particles in the cluster.

\item{}
\label{pmiddlegam}
Averaged front $<p_{(N+1)/2}^{(1)}(x;t)>$ of the central
particle (\ref{p1midgamas})
for a cluster of 10 particles
when the diffusion constants are distributed according
to (\ref{rhoDgamma}) with $\alpha\,=\,\case14$; the abscissa is the
dimensionless
variable $u=x/(4D_{0}t)^{1/2}$.

\item{}
\label{p1dbroad}
Exact averaged front $<p_{N}^{(1)}(x;t)>$ for the right particle
when the diffusion constants of the cluster are distributed according
to (\ref{rhoDbroad}) with $\mu\,=\,\case12$; the abscissa is the dimensionless
variable $u=x/(4D_{0}t)^{1/2}$. Each curve is labelled by the number
of particles in the cluster.

\end{enumerate}

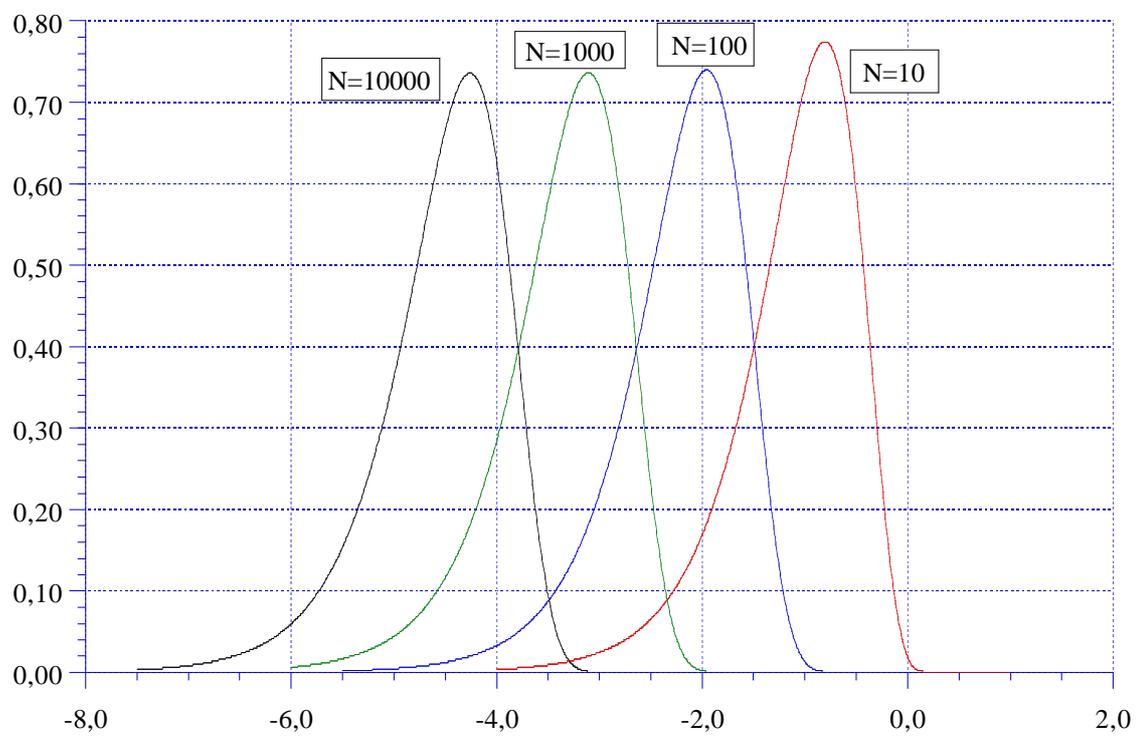
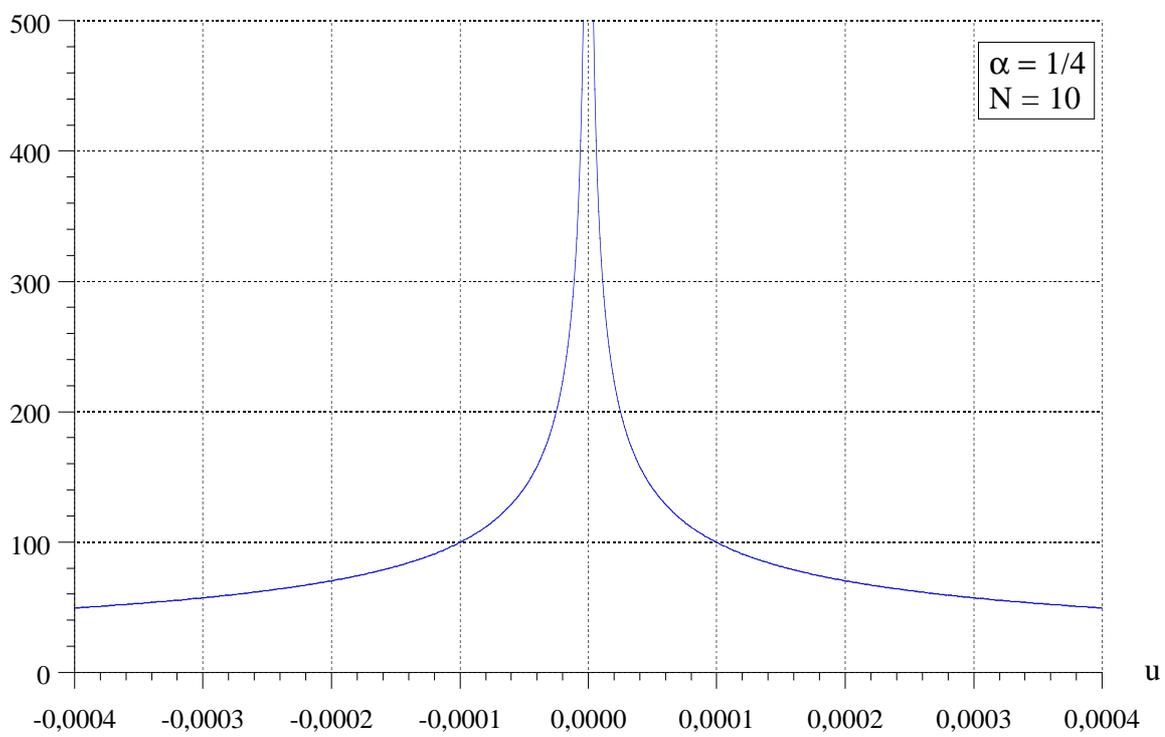
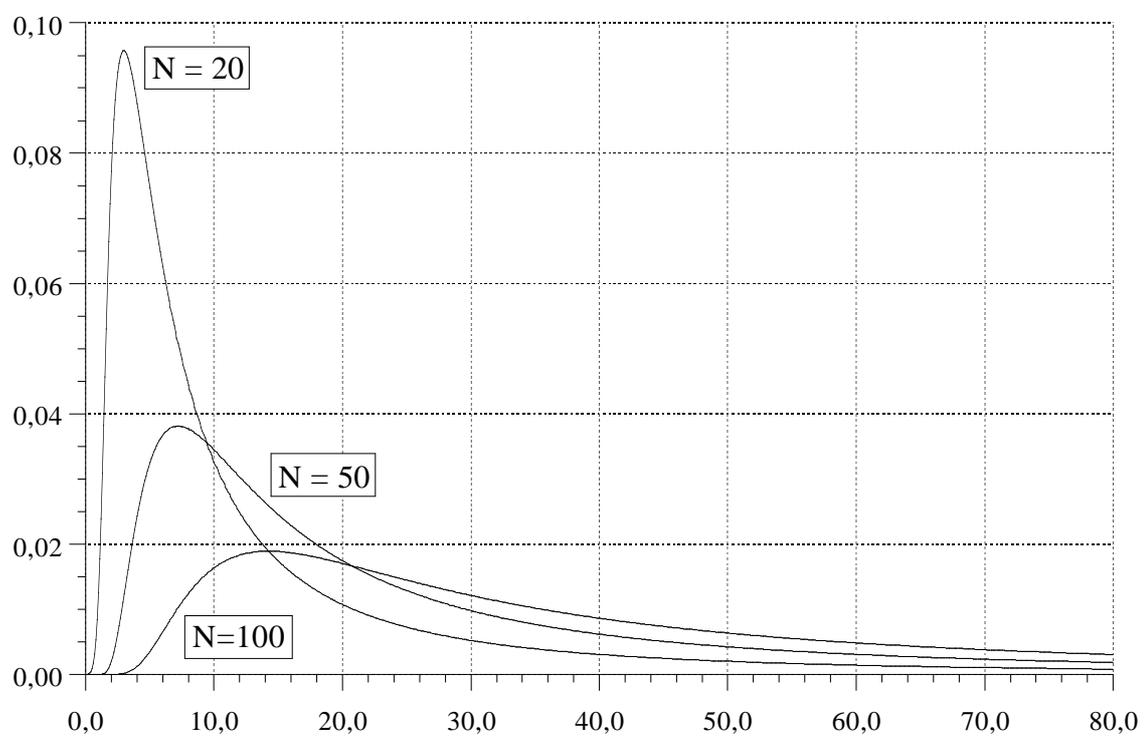
\end{document}